\documentclass{article}
\usepackage{authblk} 

\makeatletter
\renewcommand\AB@affilsepx{, \protect\Affilfont}

\makeatother

\usepackage{graphicx} 
\usepackage{dcolumn}
\usepackage{bm}
\usepackage{csquotes} 
\usepackage{acronym}
\usepackage{amsmath}
\usepackage{amsfonts}
\usepackage{cleveref}
\usepackage{url}
\usepackage{amssymb}
\usepackage{caption,subcaption}
\usepackage{csquotes}
\usepackage{appendix} 
\usepackage[sorting=none,style=numeric-comp,maxbibnames=5]{biblatex} 
\usepackage{pgfplots, pgfplotstable, filecontents}
\pgfplotsset{compat=newest}
\pgfplotsset{minor grid style = {dashed}}

\DeclareMathOperator*{\argmin}{arg\,min}

\title{QUBO-based SVM for credit card fraud detection on a real QPU}
\date{}

\author{Ettore Canonici}
\author{Filippo Caruso}
\affil{Dept.\,of Physics and Astronomy, University of Florence, via Sansone 1, 50019 Sesto Fiorentino, Italy}
\affil{European Laboratory for Non-Linear Spectroscopy (LENS), University of Florence, via Carrara 1, 50019 Sesto Fiorentino, Italy}
\affil{Email address: filippo.caruso@unifi.it}

\begin{document}

\acrodef{nisq}[NISQ]{Noisy Intermediate-Scale Quantum}
\acrodef{ml}[ML]{Machine Learning}
\acrodef{dl}[DL]{Deep Learning}
\acrodef{qml}[QML]{Quantum Machine Learning}
\acrodef{ai}[AI]{Artificial Intelligence}
\acrodef{qpu}[QPU]{Quantum Processing Unit}
\acrodef{svm}[SVM]{Support Vector Machine}
\acrodef{ann}[ANN]{Artificial Neural Network}
\acrodef{rnn}[RNN]{Recurrent Neural Network}
\acrodef{qubo}[QUBO]{Quadratic Unconstrained Binary Optimization}
\acrodef{qa}[QA]{Quantum Annealing}
\acrodef{qaa}[QAA]{Quantum Adiabatic Algorithm}
\acrodef{svm}[SVM]{Support Vector Machine}
\acrodef{ccf}[CCF]{Credit Card Fraud}
\acrodef{ccfs}[CCFs]{Credit Card Frauds}
\acrodef{qpu}[QPU]{Quantum Processing Unit}
\acrodef{qpus}[QPUs]{Quantum Processing Units}
\acrodef{aucroc}[AUC ROC]{Area Under the Curve of the Receiver Operating Characteristic}

\maketitle

\begin{abstract}
Among all the physical platforms for the realization of a \acf{qpu}, neutral atom devices are emerging as one of the main players. Their scalability, long coherence times, and the absence of manufacturing errors make them a viable candidate.. Here, we use a binary classifier model whose training is reformulated as a \acf{qubo} problem and implemented on a neutral atom \ac{qpu}. In particular, we test it on a \acf{ccf} dataset. We propose several versions of the model, including exploiting the model in ensemble learning schemes. We show that one of our proposed versions seems to achieve higher performance and lower errors, validating our claims by comparing the most popular \acf{ml} models with QUBO SVM models trained with ideal, noisy simulations and even via a real \ac{qpu}. In addition, the data obtained via real \ac{qpu} extend up to 24 atoms, confirming the model's noise robustness. We also show, by means of numerical simulations, how a certain amount of noise leads surprisingly to enhanced results. Our results represent a further step towards new quantum \ac{ml} algorithms running on neutral atom QPUs for cybersecurity applications.
\end{abstract}

\section{Introduction}    
The last decade has experienced a more and more intense research activity in the field of quantum computing as well as an increased interest of companies in its use within their businesses. Indeed, it gave rise to the experimental realization of different physical platforms for the realization of \acf{qpus}. Some of the most promising ones include: superconductors, trapped ion, spins in semiconductors, NV centers in diamond, photons, and neutral atoms trapped in optical tweezers \cite{NeutralAtoms2023, Superconducting_qc, trapped_ions_qc, spin_qubit_qc, photonic_qc, NV_centers_qc}.
However, all these systems share the fact that they are afflicted by noise introducing errors during the computation. In fact, we refer to today's quantum processor as \ac{nisq} devices~\cite{preskill2018quantum}. 
 In addition to noise, the number of performed operations performed and their duration must also be kept under control so that good results can be achieved. The number of qubits, which is relatively low, is also a problem because as their number increases, noise does increase.
For this reason, numerous studies are being carried out to find methods to reduce the effects of noise: these may be based on the addition of extra qubits to realize quantum error correction, still unfeasible today, \cite{SurfaceErrorCorrection2012, cai2021bosonic, lidar_brun_2013} or on \ac{ml} protocols \cite{RLNoiseMitigation2023, QuantumErrorMitigationWithNNs2020, MLPraticalQuantumErrorMitigation2023}. 

In this context, neutral atom devices have experienced a significant growth in interest for both the academic and non-academic community.
This intense growth culminated in the realisation of error corrected logical qubits \cite{QUERA_correction2024}, placing neutral atom devices among the most promising systems at the moment. The 48 logical qubits are obtained from 280 physical qubits (atoms) grouped into logical blocks. Leveraging this, the authors show how it is possible to perform operations while maintaining high fidelity ($99.8\%$).

In addition to this, atoms are naturally identical, hence free from manufacturing errors. Moreover, their scalability is obtained via the realization of spatially extended potentials being able to create multiple optical tweezers. 
In these devices, excited and fundamental states of atoms are used for the realization of the two (qubit) computation states. They generally benefit also from relatively long coherence times and allow working in both digital (gate mode) and analog mode (Hamiltonian mode)\cite{Henriet2020quantumcomputing}.  In the latter, one directly manipulates the Hamiltonian of the system, for instance by applying global laser pulses on the atoms. This greatly reduces the duration of quantum protocols.
Finally, neutral atom devices are naturally suited for efficiently solving combinatorial optimization problems, which makes them particularly popular in various fields: routing \cite{HybridRouting2019, QARouting2013, QCRouting2021}, scheduling \cite{QAScheduling2015,SchedulingNurse2019}, energy distribution \cite{energyDistribution2021, Silva2023Energy}, finance \cite{PasqalFinance2023, QuantumForFinanceEgger2020, corbellettoFinance2023} and also \ac{ml} applications \cite{QuantumBoltzmannMachine2018, ReverseQuantumAnnealing2019,QUBODormulations2021, QUBOSVM2020}. 
Optimization problems that can be solved naturally with such devices are the \acf{qubo} problems \cite{QUBODefinition, QUBObook}.  
Such problems, apart from being very difficult to solve exactly as they belong to the NP-hard class, find many applications in industry. Among these there are graph clustering (quantum community detection problems) \cite{Negre_2020}, traffic-flow optimization \cite{Neukart2017-er}, vehicle routing problems \cite{Feld_2019}, maximum clique problems \cite{Chapuis2019-gg},  and financial portfolio management problems \cite{Mugel2021-ym}. Therefore, many attempts have been made to find possible ways of solving them. Given their relation with the Ising model, the possibility of solving them in an approximate manner using quantum computers has been explored.
Such problems can also be solved in a classically approximate manner by simulated annealing, where temperature fluctuations are exploited to try to reach a solution to the problem. However, in some cases a quantum advantage can be obtained by exploiting quantum fluctuations and the tunnelling effect to find the minima of optimisation problems \cite{RevModPhys.90.015002}. This is the basis of adiabatic computing and quantum annealing.
For such problems, \ac{qaa}, i.e. a process in which a quantum Hamiltonian is evolved very slowly into the one representing the desired problem, can be exploited. The state of the system, in turn, changes from the ground state of the initial Hamiltonian to the one of the target Hamiltonian, thus obtaining the solution of the problem.
Potentially, the high connectivity and flexibility of atom topology seem to make them more advantageous than other quantum annealing platforms \cite{D-wave}. 

Since it is possible to reformulate the training process of a \ac{ml} model as a \ac{qubo} problem \cite{QUBOObjectFunction}, a neutral atom device can be used to train \ac{ml} models via \ac{qaa}. Specifically, the model that will be trained is a QUBO-based version of \ac{svm} \cite{QuboSVM2019}, which will be called QUBO SVM from here on.

Supervised Learning, as a branch of ML, is characterised by the presence of ground truths (labels) that are associated with the data. These ground truths play a crucial role in the training and testing phase of the model. For this purpose, the dataset is divided into two parts (at least), one for training and one for performance evaluation, i.e. training and test sets, respectively. Within supervised learning, it is possible to identify classification and regression tasks. The former concerns the training of a model capable of assigning data to two (or more) classes.  In the latter, a model is trained to predict a continuous value of one or more parameters.
\ac{svm}s are a family of widely used classifier models that are appreciated for their stability \cite{HastieTheElements2009, shalev2014understanding}, meaning that small differences in the training set do not cause significant changes in the results.
In general, these models are used when the dataset is small, but there are applications where they are used on top of neural networks with significant gains in performance \cite{DNNSVMkim, DNNSVMAtm, ZAREAPOOR20184}.
In particular, \ac{svm} versions based on digital quantum computing have been developed \cite{gateQSVM} which, however, suffer from the limited number of qubits, noise, and the fact that the quantum processors must be used for both training and test set.

This makes them unusable on large datasets because of the cost and access limitations of \ac{qpus}.
In this manuscripts, we will compare the performance of the QUBO SVM models trained via \ac{qaa} on a dataset of particular relevance: \ac{ccf}, i.e., a scenario in which someone other than the owners makes an unlawful transaction using a credit card or account details. The problem is of crucial relevance given the exponential growth of electronic payments, resulting in annual losses of billions of dollars.
Automated detection of \ac{ccf}s is therefore a hot topic of research \cite{CCFDL2022}. 
This problem has already been addressed in an unsupervised manner by Ref. \cite{Mellaerts2023QuantuminspiredAD}, however, this approach currently restricts its applicability because it cannot be implemented on current \ac{qpus} due to its unsupervised nature. In fact, it is necessary to provide the entire dataset to the \ac{qpu} by encoding it into a single \ac{qubo} matrix, whose size scales with the number of qubits. Although the model shows very good performance, current \ac{qpus} do not have sufficiently enough number of qubits for application-relevant data sets. 
Our approach, on the other hand, requires that only the training part exploits a \ac{qpu}, while the testing phase exploits classical hardware. Therefore, since the QUBO matrix is obtained only from the training data, it is sufficient that the size of the training set is chosen in such a way as to respect the limits of today's \ac{qpus}. Thus, what is proposed in this manuscript is applicable on today's \ac{nisq} devices.

The QUBO SVM model is first implemented and simulated on classical hardware, then tested on a real QPU by extending the number of qubits (or similarly atoms) used. For the simulation on classical hardware, we exploit the \emph{Python} library \emph{Pulser} \cite{silverio2022pulser}, which allows ideal and noisy simulations. 
\emph{Pulser} is developed by \emph{Pasqal} \cite{pasqal}, a company that builds \ac{qpus} based on neutral atoms, which, in some versions, can leverage more than 100 atoms. Via \emph{Pulser}, it is possible to run simulations either on real machines or on classical hardware.
As far as the implementation on \ac{qpus} is concerned, the real Fresnel neutral atom \ac{qpu} developed by \emph{Pasqal} is used, which, in the experimental configuration available at the time of the experiment, can use up to 25 atoms.

This manuscript is organised as follows: in Sec. \ref{sec:qubo problem}, a brief introduction to \ac{qubo} problems is given, and in Sec. \ref{sec:CCF detection}, a quick overview of the phenomenon of \ac{ccfs} and their detection is given. The sections \ref{sec:SVM} and \ref{sec:qubo svm} respectively introduce the \ac{svm} model with some of its applications and modified versions. In Sec. \ref{sec:dataset}, the used \ac{ccf} dataset is presented and described, while Sec. \ref{sec:qubo svm} discusses the implementation of the \ac{qubo} problem concerning the training of an \ac{svm} model on a neutral atom \ac{qpu}. Finally, in Secs. \ref{sec:results} and \ref{sec:conclusion}, more general comments on our results are presented and some final conclusions and outlooks are drawn.

\section{QUBO problem}\label{sec:qubo problem}
\ac{qubo} problems are combinatorial optimization problems of particular importance for adiabatic quantum computing given the connection with Ising models. They fall among the NP-hard problems, and therefore alternative ways to the classical ones are being explored for their solution. Quantum computers have shown advantages in their approximate solution. In particular, in adiabatic quantum computing \ac{qubo} problems are solved by \ac{qaa}.
Formally, a \ac{qubo} problem is defined as the minimization of the cost function
\begin{equation}\label{QUBO definition}
    E = a^{\top} Q a = \sum_{i,j=1}^{n} a_i Q_{ij} a_j,
\end{equation}
where $a \in\mathbb{B}^{n}$ is a binary vector, $\mathbb{B} = \{ 0,1\}$ is the set of binary variables and $Q \in \mathbb{R}^{n \times n}$ is known as \ac{qubo} weight matrix.
The goal of the minimization is to find a binary vector $a^*$ such that
\begin{equation}
    a^* = \argmin_{a \in\mathbb{B}^{n}} E .
\end{equation}

\section{\ac{ccf} detection}\label{sec:CCF detection}
Recent years have seen a growth in electronic payments based on credit cards, both in physical shops and in online payments. With them, the number of identity thefts is also on the rise. This poses a major new problem, as \ac{ccfs} are responsible for financial losses for credit card holders and credit institutions themselves. Confirming this, in 2018 the value of fraudulent transactions related to cards issued in the eurozone alone amounted to approximately 1.8 billion euros \cite{ECB2020}. In 2025, worldwide fraud-related losses are expected to reach approximately 35 billion dollars \cite{Nilson2020}. For this reason, financial institutions are prioritising the development of algorithms capable of detecting and preventing losses due to such activities. For this reason, financial institutions are prioritising the development of algorithms capable of detecting and preventing losses due to such activities. In particular, \ac{ml} and \ac{dl} algorithms are being applied to decide whether an incoming transaction is related to a fraud attempt or not.
However, given its importance, it is a very active field of research, with constant improvements and in which many techniques from different backgrounds are applied.

\section{SVM}\label{sec:SVM}
Binary classification is one of the most famous and common \ac{ml} tasks: it involves training a parametric model that becomes able to assign data to two distinct categories via supervised training. 
\ac{svm} is a very famous classical \ac{ml} model used to perform binary classifications. 
With mapping data in high-dimensionality spaces, it is possible to classify both linearly separable data and non-linearly separable data. However, this results in computational costs that can grow exponentially when dealing with large amounts of data. This problem can be circumvented by using kernels. Indeed, when data appears in the form of a scalar product, instead of mapping the data into another space and calculating their scalar product, it is possible to replace it directly with the inner product of the data in the new space. The use of kernels leads to an overall less computationally expensive problem and has the advantage of strong mathematical foundations.

In the field of supervised learning, it is well known that \ac{svm} can offer good performance even with small datasets, unlike neural networks requiring a lot of data to be trained. We now consider a dataset 
\begin{equation}
    D = \{ (x_n, y_n) : x_n \in\mathbb{R}^{d}, y_n = \pm 1 \}_{n=0, \dots, M-1}, 
\end{equation}
where $x_n$ is a sample (i.e. the feature vector) and $y_n$ the associated target. We call the classes $y_n \in \{1,-1\}$ $\forall n$ respectively as \textquote{positive} and \textquote{negative}. The reader interested in more technical details of the \ac{svm} model can find them in the appendix \ref{appendix:SVM}.

\section{QUBO Support Vector Classifier}\label{sec:qubo svm}
After introducing the \ac{svm} model and the \ac{qubo} problem we can show that is possible to reformulate the training process of a \ac{svm} model into a \ac{qubo} optimization problem. From now on we will refer to this model as QUBO SVM. The reformulation involves rewriting Eq. \ref{svm training} in a form similar to Eq. \ref{QUBO definition}. At the end of some manipulations, it is possible to obtain the following \ac{qubo} matrix:
\begin{equation}\label{qubo matrix par}
    Q_{Kn+k, Km+j} = \frac{1}{2}B^{k+j} y_n y_m (k(x_n, x_m) + \xi) - \delta_{nm}\delta_{kj}B^k,
\end{equation}
where $B$, $K$ and $\xi$ are constants, $k(\cdot)$ is a kernel function and $\delta_ij$ is Kronecker's delta.
It can be observed that the square matrix $Q$ has a linear dimension equal to $K \times N$, where $N$ is the number of samples in the training set. This is related to the fact that the encoding of continuous data in the form of binary sequences requires their discretization. In our case, the encoding is done via the Eq. \ref{encoding}.
For the interested reader, the full derivation can be found in appendix \ref{appendix:QUBO SVM} . 

\begin{figure}[h!]
    \centering
\includegraphics[width=0.9\textwidth]{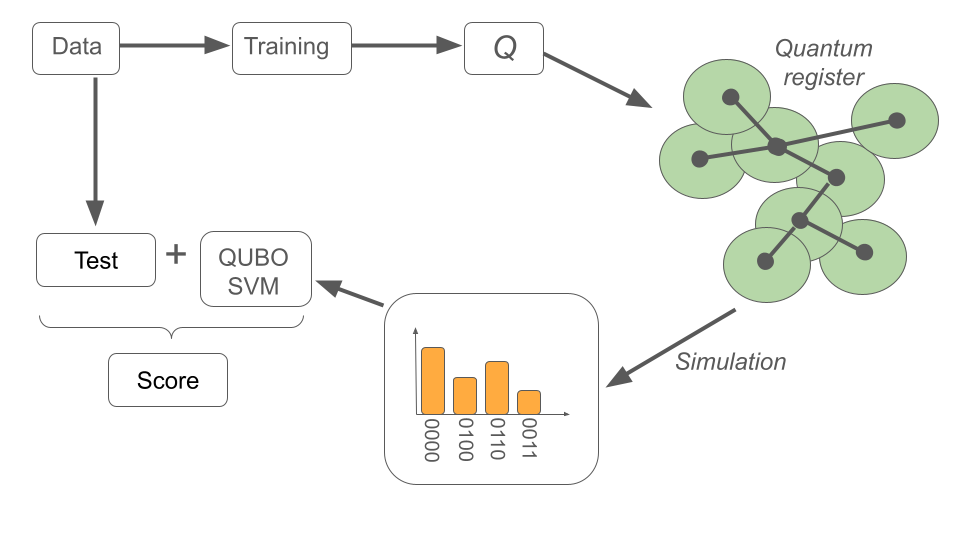}
    \caption{Schematic representation of our protocol.}
    \label{fig:protocol}
\end{figure}

As previously mentioned, each bitstring obtained as output from the quantum device measurement can be used as the decision hyperplane of a \ac{svm} model. Hence, there is a close connection between the matrix $Q$ and the objective function to be minimised. We do not actually use $Q$ directly, we modify the position of the in such a way as to replicate $Q$ as closely as possible in the Hamiltonian $H_Q$. Since this optimisation is the training of a \ac{svm} model, finding a solution means finding the best \ac{svm} model for the specified training set (hence for that given matrix $Q$). Since we obtain a distribution of solutions (bitstrings resulting from the measurement process), we have at our disposal a distribution of \ac{svm} models that we can potentially aggregate to obtain performance gains. The aforementioned protocol is schematised in the Fig. \ref{fig:protocol}.

\section{Dataset and metrics}\label{sec:dataset}
The \ac{ccf} detection is a binary classification with an unbalanced dataset, namely a supervised anomaly detection. By definition, anomalies are rare events that have great impact. The considered dataset concerns \ac{ccf} detection, found on Kaggle \cite{creditcardfraudDataset_2016}. 
We chose to search for a dataset on this website since there are many datasets of applicational relevance, often companies organise challenges on kaggle offering monatary rewards to the winners. The \ac{ccf} detection dataset was chosen because, among the datasets present, it represents a problem of considerable relevance. It has a very marked imbalance between classes: 284315 \textquote{normal} transactions and 492 \textquote{anomalous} transactions (about $0.17\%$ of transactions are the positives). In addition, it contains only numerical input variables which are the result of a PCA transformation. Due to confidentiality issues, the original features and more background information about the data are not provided.

Specifically, we divide the dataset into two equal parts. A part is then used as a test set, so it is ignored until that point. Instead, the other part is divided into a validation set and a training set. The latter contains 6 training samples. These training and validation sets are first resampled, and then used for training and validation procedures. We repeat this process 10 times, collecting metrics and providing estimates with associated errors in the form of mean values and standard deviations calculated across the 10 analyses on different datasets.
In addition, to provide performance and error scaling, we repeat the analysis using training sets of 7 and 8 training samples.

Since this is an unbalanced classification problem in which the focus is on the minority class, accuracy is not a good metric to use. Recall and balanced accuracy are used instead:
\begin{equation}
    \text{recall} = \frac{True\text{ }Positives}{All\text{ }Positives} = \frac{True\text{ }Positives}{True\text{ }Positives + False\text{ }Negatives},
\end{equation}
\begin{equation}
\begin{split}
    \text{balanced accuracy} &= \frac{1}{2} \left[ \frac{True\text{ }Positives}{True\text{ }Positives + False\text{ }Negatives} + \right. \\
    &\left. \frac{True\text{ }Negatives}{True\text{ }Negatives + False\text{ }Positives} \right].
\end{split}
\end{equation}
Recall is the fraction of relevant instances that are retrieved by the model, while balanced accuracy is a metric specifically developed to generalize accuracy in the case of unbalanced datasets.
A high recall means that the number of false negatives is low, i.e., the number of frauds that the model fails to recognize is low compared to those that it does recognize.
The number of false positives (when the model classifies a transaction as fraud by mistaking it) is not as important as false negatives, since with subsequent checks it can be determined whether it was fraud or a false alarm (e.g., by sending verification messages on the credit card owner's smartphone, etc.). This is because the scenario where the bank loses money is the false negative case.
In this scenario, with such an imbalance of classes, the application of resampling algorithms is one of the most adopted practices. Initially, the SMOTE oversampling algorithm \cite{SMOTE} is applied to increase the minority class data (the frauds), then random undersampling \cite{IMBLEARN} is used to reduce the size of the dataset to balance the classification problem. At the end, a balanced dataset containing 500 samples is obtained. This will then be divided into training and validation sets.
The test set, instead, has not been rebalanced and has the same class ratio as the original data set ($0.17\%$ fraud). This is because the test set only needs to simulate a scenario that is as realistic as possible, unlike the other splits that can instead be modified to attempts making the model to learn better. Treating the test set separately avoids data leakage that would lead to overestimation of performance.

To be sure about the performance of the QUBO SVM models, they are compared with the main classical models in the literature: classical \ac{svm} and Decision Tree. In addition, various versions of the QUBO SVM models are trained using ideal and noise simulations (denoted respectively with the letter \textquote{i} and \textquote{N}). Furthermore, two types of models are tested: QUBO SVM and a stacked ensemble configuration in which the QUBO SVM model is used as a meta-model, i.e. it is trained using the outputs of other models as data.
After ensuring that the models actually work, both configurations are tested on a real QPU, exploring regimes that are not easily simulated with local clusters.

The reader interested in all the information on the various model configurations (including the number of measurements used for each training) that have been used can find it in the appendix \ref{appendix:models}.

\section{Neutral atom implementation}\label{sec:implementation}
We now describe the model implementation on a neutral atom \ac{nisq} device. We distinguish three phases: encoding, training and testing. Let us point out that only the second phase takes place on a \ac{qpu}, while the first and third ones are pre- and post-processing via traditional CPU.
Encoding is achieved by computing the \ac{qubo} matrix of Eq. \eqref{qubo matrix}: this is done by specifying all the training data together. This means that the linear dimension of the (square) matrix will be $K \times N$. Hence, the parameter $K$ plays an important role in terms of the number of qubits required to implement the algorithm. $K \times N$ are the required atoms. The encoding in this case takes place on the quantum register. 

We recall that for a \ac{qpu} of neutral atoms the (global) Hamiltonian has usually the form
\begin{equation}
    H = \frac{\hslash \Omega(t)}{2} \sum_{i} \sigma_{i}^{x} - \frac{\hslash \delta(t)}{2} \sum_{i} \sigma_{i}^{z} + \sum_{i<j} U_{ij}n_{i}n_{j}.
\end{equation}
Here, $\Omega$ and $\delta$ are, respectively, the laser amplitude and the detuning, $n_i = (1 + \sigma_{i}^{z})/2$ and with $\sigma_{i}^{z}$ representing the component along the $z$-axis of the pauli vector applied to the $i$-th atom.
By tuning the positions of the atoms, the value of $\Omega$ and $\delta$, we can make $H$ as close as possible to $Q$.
This is done using the COBYLA optimizer \cite{cobyla1992} and imposing the physical constraints of relative distance between atoms that are feasible on the specific real \ac{qpu}. As far as the \ac{qpu} is concerned, due to the prototype available at the time of the simulations, the atoms in the quantum register can only be arranged on the vertices of a triangular lattice with a fixed lattice constant. Therefore, now the co-ordinate optimisation no longer takes place on continuous but discrete space, which is why now the best spatial configuration of atoms on a triangular lattice is found by simulated annealing. Although this appears to be a more disadvantageous condition for satisfying $H_Q \simeq Q$, we will show that from the results it does not appear that the model suffers from such a limiting constraint. 

\begin{figure}[h!]
    \centering
\includegraphics[width=1.\textwidth]{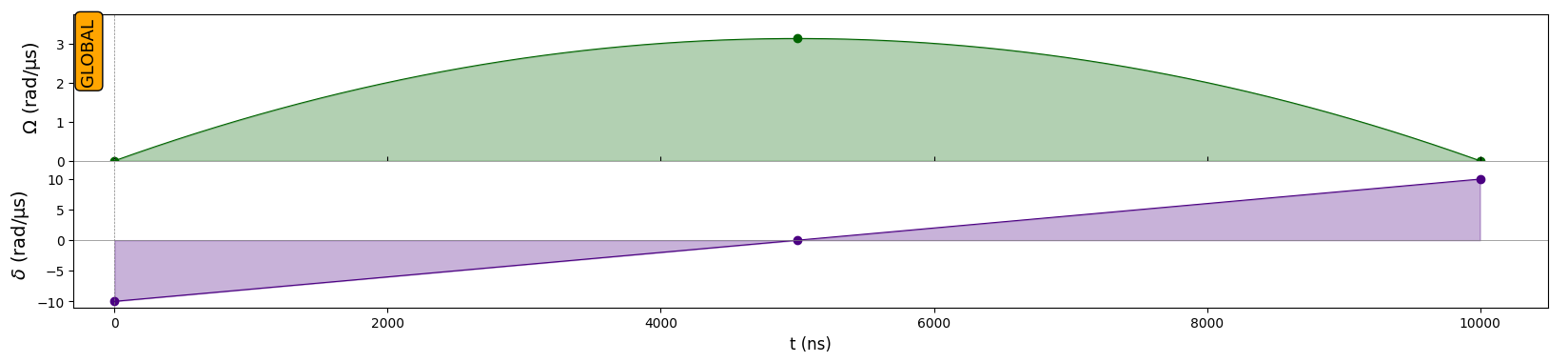}
    \caption{pulse sequence of the protocol, the the height of the smooth green curve depends specifically on the \ac{qubo} matrix.}
    \label{fig:pulse sequence}
\end{figure}

Once the encoding is done, the desired quantum register topology is realized on the \ac{qpu}, the pulse is applied, and the final measurement of the entire quantum system is taken.
The applied control consists of a pulse sequence with each pulse of duration equal to 10 $\mu s$ and maximum amplitude equal to
\begin{equation}
\Omega=
    \begin{cases}
    median(Q),& \text{if }  median(Q)< \Omega_{max}\\
    \Omega_{max},              & \text{otherwise},
\end{cases}
\end{equation}
where $\Omega_{max}=15.71$ $rad/\mu s$ is the maximum amplitude being possible on the \emph{Pasqal} device. At the same time, a linear detuning ramp is applied, starting at -10 $rad/\mu s$ and ending at 10 $rad/\mu s$ after 10 $\mu s$. The obatined sequence is shown in Fig. \ref{fig:pulse sequence}. 
The sequence applied in the real quantum device is the same, exception made for the duration that is reduced to 5 $\mu s$.

We denote with $H_Q$ the Hamiltonian obtained after tuning the coordinates and fixing the sequence. In any case, we recall the goal of this procedure is to obtain an Hamiltonian such that $H_Q \simeq Q$.
Finally, we repeat the preparation-pulse-readout cycle $N_{shots}$ times. This yields a distribution of atomic states with the corresponding probabilities: if the problem has been well encoded in the Hamiltonian, the highest probability states are the best solutions of the \ac{qubo} problem.
But the problem framed as a \ac{qubo} is the training of an \ac{svm} model, i.e. finding a hyperplane that best separates the points of the two classes while committing as few errors as possible.
Therefore, each quantum state obtained (a string of binary variables) is a possible decision boundary defined by $\alpha_n$.
The state with the largest probability is the best solution to the \ac{qubo} problem and it contains the coefficients of the best model. 
Since a particular \ac{svm} model is completely defined by the hyperplane (which is defined by the coefficients $\alpha$), to test a model it is sufficient to use a particular $\alpha_n$ in the decision function \eqref{decision function} and provide new data as input to the model.
This testing operation is done on a classical computer. At this point, the predicted class can be derived from the decision function. By repeating this it is possible to compute the predictions on a whole test set. After that, one can use the metrics for classification used in classical models: accuracy, precision, recall, f1-score, balanced accuracy, \ac{aucroc} and so on.

However, this classifier allows us to experiment different approaches towards the best performing one: (i) using the trained model as it is or (ii) using it as a base model in different kinds of ensemble techniques. The latter is a ML technique that combines the predictions of several trained models together to form a new model that exhibits superior performance, lower bias, robustness, etc.
There are various ways of combining models, here we exploit average voting and stacking. In the first case, many models are trained on the same training set and subsequently tested on the same test set. The final predictions are an average of the predictions of the individual models. In stacking, on the other hand, two layers of models are used. In the first layer, several models are trained on the same training set and then tested on the same test set. In this case, however, the outputs of the models in the first layer are used to train and test the model in the second layer, also known as the meta-model. These types of approaches are exploited in \ac{ccf} detection \cite{EnsembleLOUZADA201211583, EnsembleSUNDARKUMAR2015368}. More information on ensemble techniques can be found in the appendix \ref{appendix:ensemble} and in Ref.\cite{HastieTheElements2009}.

In particular, since a full training of $N_{shots}$ measurements leads to a distribution of states with the relative probabilities, it is possible to set up a full quantum average voting strategy. In fact, remembering that each observed bitstring represents a possible solution of the \ac{qubo} problem), one can potentially obtain up to $2^{N_{atoms}}$ states (and thus solutions of the training problem). 
It is interested to see if aggregating multiple models into a single model yields more robust predictions.
Then, operationally, the predictions on the test set are computed using each bitstring (state) obtained from the training process. The predictions obtained from the various models are then aggregated by taking an average.
In addition, by providing a validation set and specifying with respect to which metric one wants to optimize, it is possible to repeat this procedure by finding the number of aggregate models maximizing that metric (one starts with the highest probability states eventually trying to use all of them). At the end of this procedure, one obtains the classifier ensemble composed of the models that maximize the specified metric.

Furthermore, it is possible to use the proposed \ac{qubo} SVM model as a meta-model in a hybrid quantum-classical stacked ensemble learning procedure, in which the quantum model is trained using the outputs of classical models as input.

\section{Results and discussion} \label{sec:results}
\begin{figure}
  \centering
\begin{subfigure}{.75\textwidth}
  \includegraphics[width=1.05\linewidth]{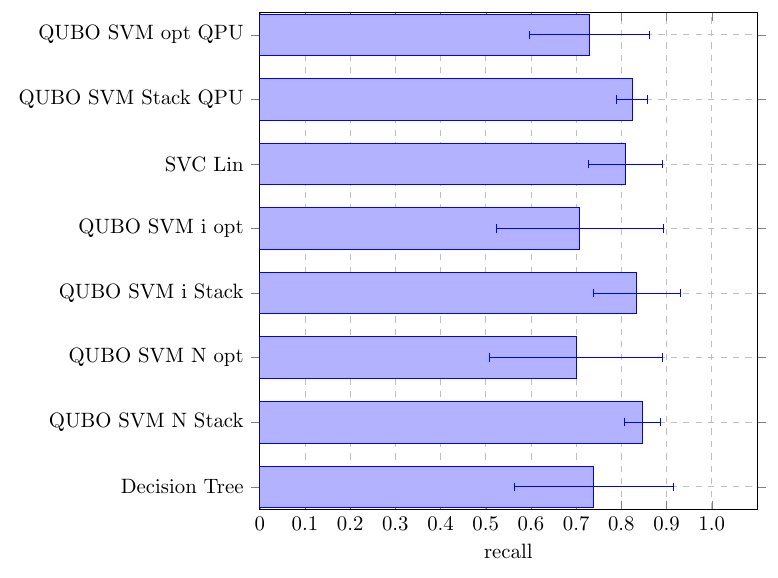}  
  \caption{\centering 8 training samples, 16 qubits simulations.}
  \label{fig:recall8_new}
\end{subfigure}
\newline
\centering
\begin{subfigure}{.75\textwidth}
  \includegraphics[width=1\linewidth]{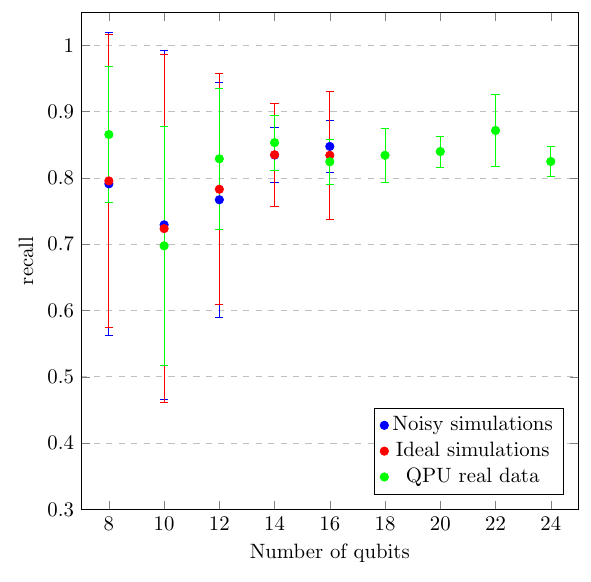}  
  \caption{\centering Recall scaling with the number of qubits.}
  \label{fig:trendline_recall_new}
\end{subfigure}
\caption{Comparison of recall between classical and QUBO SVM models using 8 training samples (16 qubits),  subfigure (a). Subfigure (b) shows the recall scaling in both ideal, noisy and QPU QUBO SVM stack models. In all subfigures, data with error are intended as mean and standard deviation calculated on 10 different splittings of the dataset. For the complete list of the used models see Tab. \ref{tab:models}.}
\label{fig:Recall_new}
\end{figure}

\begin{figure}
  \centering
\begin{subfigure}{.75\textwidth}
  \includegraphics[width=1.05\linewidth]{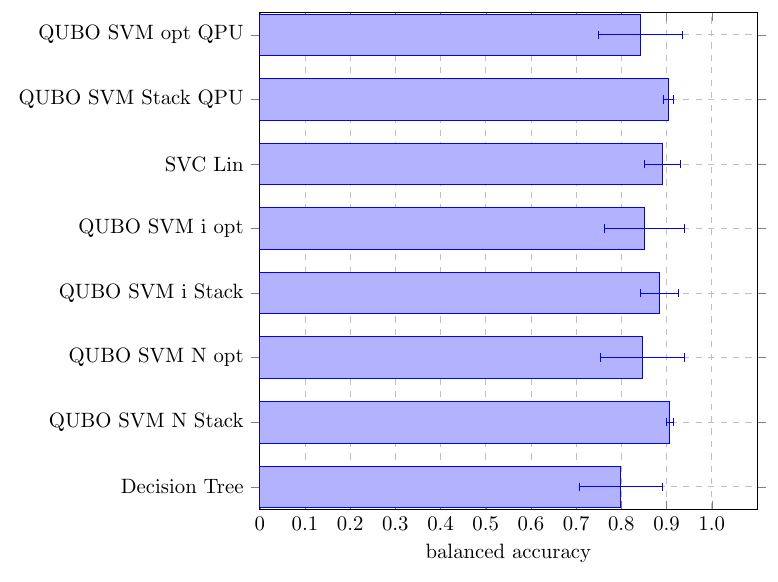}  
  \caption{\centering 8 training samples, 16 qubits simulations.}
  \label{fig:baò_acc8_new}
\end{subfigure}
\newline
\centering
\begin{subfigure}{.75\textwidth}
  \includegraphics[width=1.\linewidth]{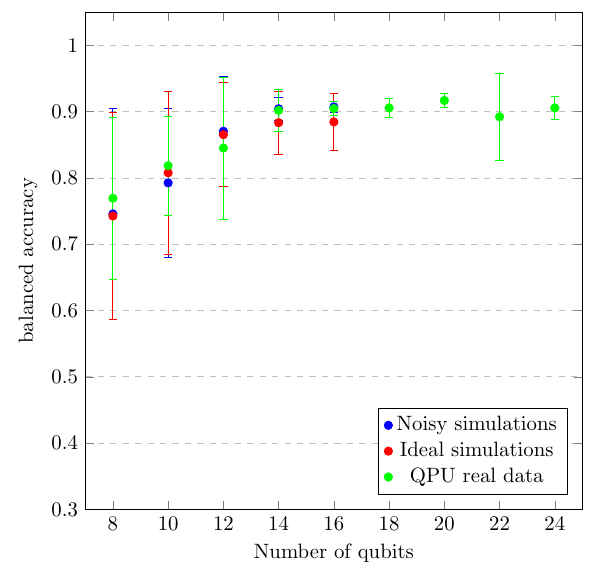}  
  \caption{\centering Balanced accuracy scaling with the number of qubits.}
  \label{fig:trendline_bal_acc_new}
\end{subfigure}
\caption{Balanced accuracy analysis using 4, 5, 6, 7, and 8 training samples, subfigure (a). Subfigure (b) shows the recall scaling in both ideal, noisy and QPU QUBO SVM stack models. In all subfigures, data with error are intended as mean and standard deviation calculated on 10 different splittings of the dataset. For the complete list of the used models see Tab. \ref{tab:models}.}
\label{fig:Bal_acc_new}
\end{figure}

In this extreme scenario, all the QUBO SVM models seem to work well (as compared to the classical models) and the error bars are about the same as for the other models, as shown in Figs. \ref{fig:Recall_new} and \ref{fig:Bal_acc_new}. Although \ac{svm}-based models do not perform well with unbalanced datasets, we have mitigated this problem with resampling techniques. 

\begin{figure}
\begin{subfigure}{.5\textwidth}
  \centering
  \includegraphics[width=0.99\linewidth, height=4.6cm]{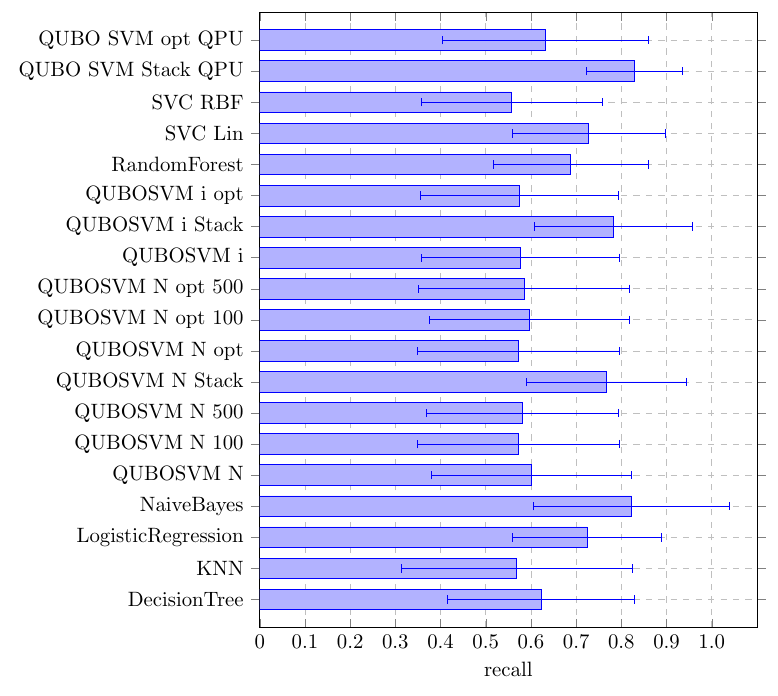} 
  \caption{\centering 6 training samples, simulations with 12 qubits.}
  \label{fig:recall6}
\end{subfigure}
\begin{subfigure}{.5\textwidth}
  \centering
  \includegraphics[width=0.99\linewidth, height=4.6cm]{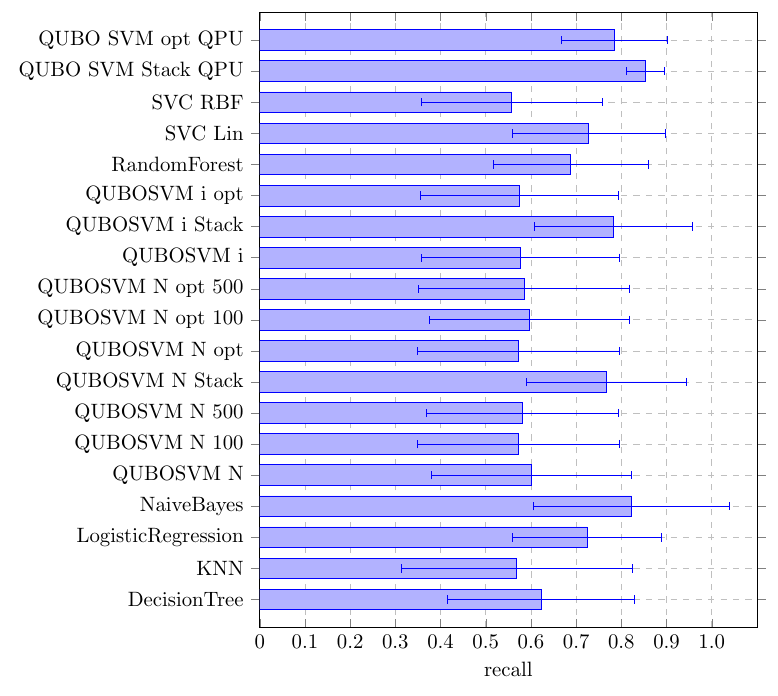}
  \caption{\centering 7 training samples, simulations with 14 qubits.}
  \label{fig:recall7}
\end{subfigure}
\newline
\begin{subfigure}{.5\textwidth}
  \centering
  \includegraphics[width=0.99\linewidth, height=4.6cm]{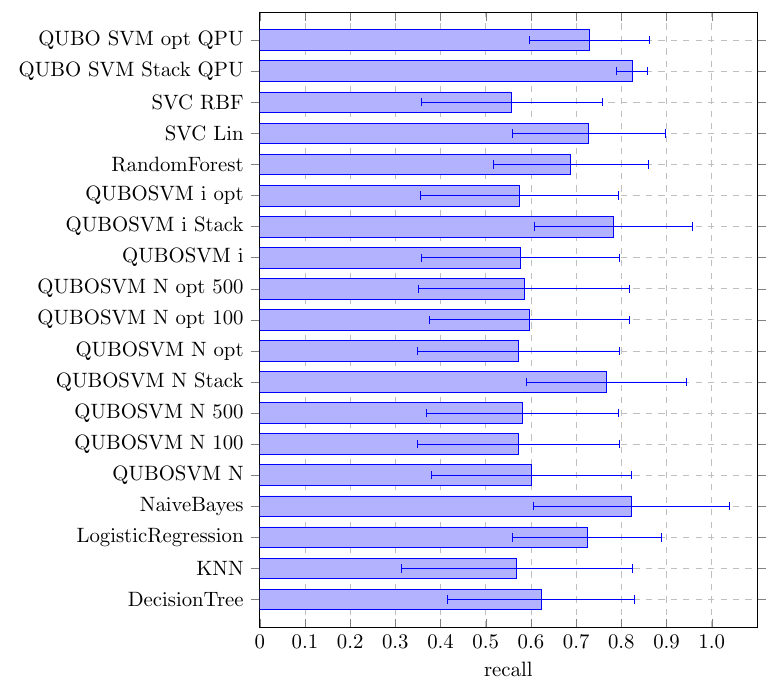} 
  \caption{\centering 8 training samples, simulations with 16 qubits.}
  \label{fig:recall8}
\end{subfigure}
\begin{subfigure}{.5\textwidth}
  \centering
  \includegraphics[width=0.99\linewidth, height=4.6cm]{trendline_recall_QPU.pdf} 
  \caption{\centering Scaling of recall in stack configuration of QUBO SVM.}
  \label{fig:recall scaling}
\end{subfigure}
\caption{Recall analysis using 6, 7, and 8 training samples. Figure (a), (b) and (c), respectively. Subfigure (d) shows the recall scaling in both ideal, noisy and \ac{qpu} QUBO SVM stack models. The \ac{qpu} version also contains simulations with 18, 20, 22 and 24 qubits.  In all subfigures, data with error are intended as mean and standard deviation calculated on 10 different splittings of the dataset. For the complete list of the used models see Tab. \ref{tab:models}.}
\label{fig:Recall}
\end{figure}

\begin{figure}
\begin{subfigure}{.5\textwidth}
  \centering
  \includegraphics[width=0.99\linewidth, height=4.6cm]{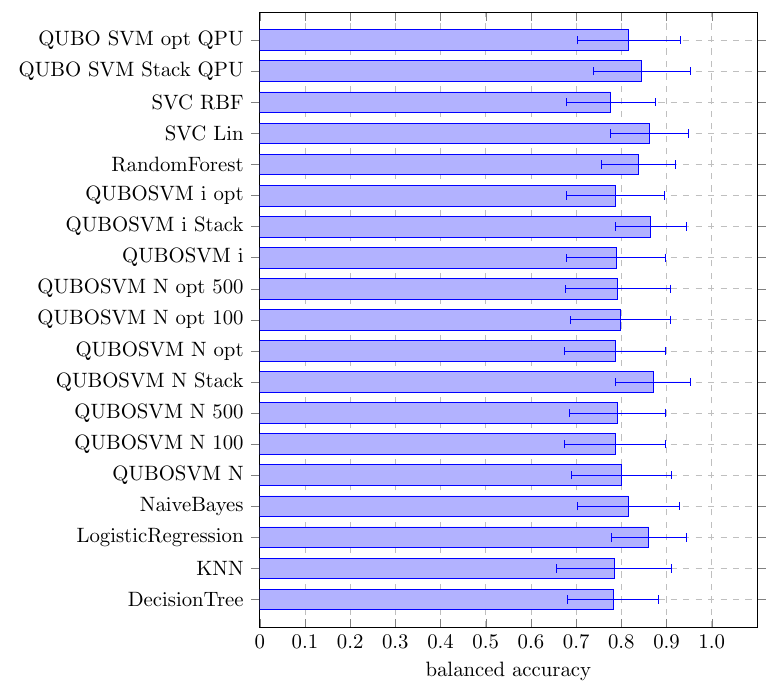}  
  \caption{\centering 6 training samples, simulations with 12 qubits.}
  \label{fig:balacc6}
\end{subfigure}
\begin{subfigure}{.5\textwidth}
  \centering
  \includegraphics[width=0.99\linewidth, height=4.6cm]{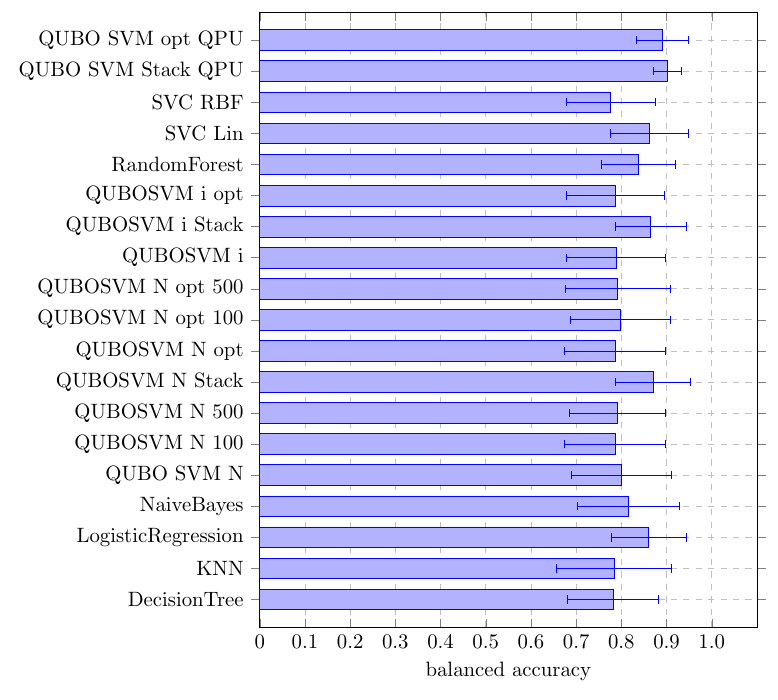}  
  \caption{\centering 7 training samples, simulations with 14 qubits.}
  \label{fig:balacc7}
\end{subfigure}
\newline
\begin{subfigure}{.5\textwidth}
  \centering
  \includegraphics[width=0.99\linewidth, height=4.6cm]{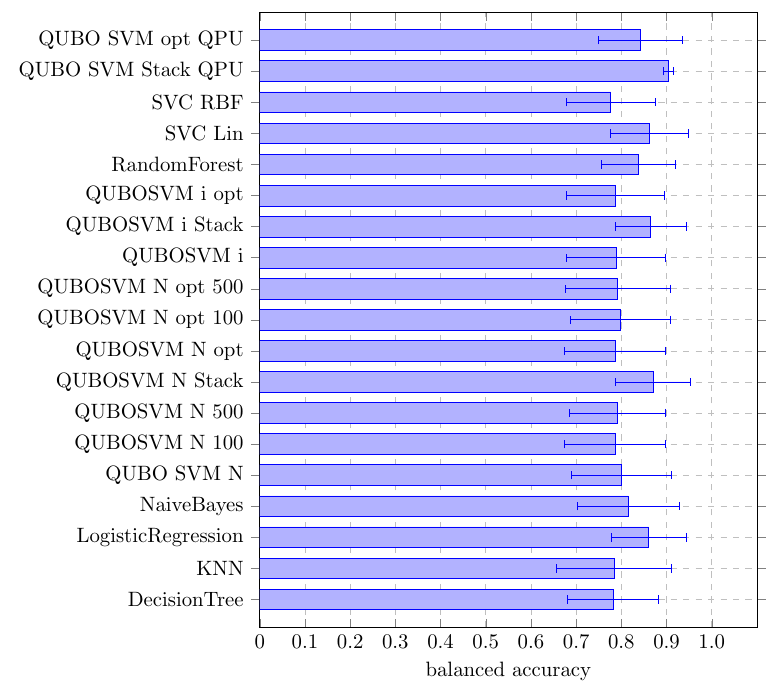}  
  \caption{\centering 8 training samples, simulations with 16 qubits.}
  \label{fig:balacc8}
\end{subfigure}
\begin{subfigure}{.5\textwidth}
  \centering
  \includegraphics[width=0.99\linewidth, height=4.6cm]{trendline_BalAcc_QPU.pdf}  
  \caption{\centering Scaling of balanced accuracy in stack configuration.}
  \label{fig:balacc scaling}
\end{subfigure}
\caption{Balanced accuracy analysis using 6, 7, and 8 training samples. Figure (a), (b) and (c), respectively. Subfigure (d) shows the balanced accuracy scaling in both ideal, noisy and \ac{qpu} QUBO SVM stack models. Again, the \ac{qpu} version contains data obtained with 18, 20, 20 24 qubits simulations.} In all subfigures, data with error are intended as mean and standard deviation calculated on 10 different splittings of the dataset. For the complete list of the used models see Tab. \ref{tab:models}.
\label{fig:BalAcc}
\end{figure}

From Fig. \ref{fig:Recall} and \ref{fig:BalAcc} it can be concluded that there appears to be no substantial difference in performance using different numbers of shots: respectively 1000, 500 and 100. Furthermore, the training simulations count on average $76 (\pm 5)$ shots each. Therefore, there is evidence for the models working even with only a few shots. This results in substantial savings in \ac{qpu} utilization, a reduction in execution time and related costs for the end user. In addition, there does not seem to be much difference between the optimized and non-optimized versions, probably because the models that perform better than random guessing are very similar to each other. Another observation is that the model seems to be rather robust to noise (both simulated and real); in fact we do not see large differences between the performance of each model in the ideal, noisy and QPU versions. Finally, the pipeline in stacked configuration seems to provide very interesting results both for the high performance, when referred to the examined models, and for the lower standard deviation with respect to all models. 
It can also be seen that performance increases as the size of the training set increases, this is interesting because for each additional sample two atoms are used. Therefore, noise does not seem to play such a negative role in the performance of the model. This is true both for simulated data with noise and for data from a real QPU, where noise is certainly present. It is an important result, because it shows that it is possible to make \ac{qml} algorithms that can really be implemented on actual \ac{qpus} to make predictions on real datasets of relevance.

Figures \ref{fig:trendline_recall_new} and \ref{fig:trendline_bal_acc_new} compare the performance of stack configuration models in the three versions: ideal, with noise and \ac{qpu}. In particular, while the versions simulated with classical hardware are tested with 8, 10, 12, 14, 16 atoms respectively, the versions trained with \ac{qpu} are also tested in the following scenarios: 18, 20, 22 and 24 atoms. We can see that, as the training set increases, the performance increases, while the standard deviation decreases. This applies to all models, which show no appreciable performance differences. Surprisingly, there seems to be a greater reduction in standard deviation in the noisy model, especially for the models trained with the \ac{qpu}.
In order to validate this intuition quantitatively, we study the scalability of performance when varying noise parameters: laser intensity fluctuations, state preparation and measurement errors, temperature of the atoms in the quantum register and laser width. The models whose scaling is evaluated are the ones that are additionally trained on the \ac{qpu}, i.e. the QUBO SVM model in stack configuration and the one trained through 100 shots only and optimised (as described above).

\begin{figure}[ht]
\begin{subfigure}{.5\textwidth}
  \centering
  \includegraphics[width=1.0\linewidth]{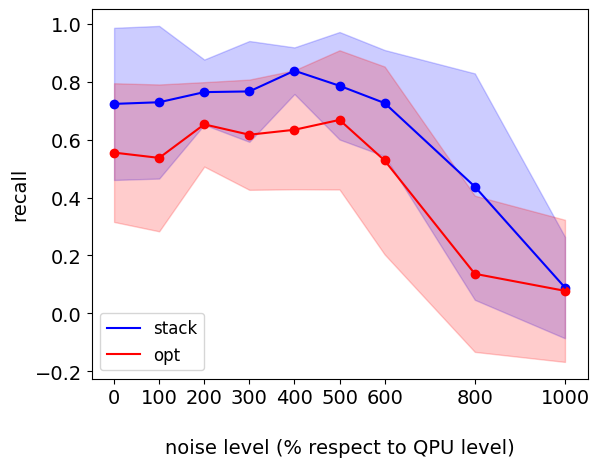}  
  \caption{Recall}
  \label{fig:sub-first}
\end{subfigure}
\begin{subfigure}{.5\textwidth}
  \centering
  \includegraphics[width=1.0\linewidth]{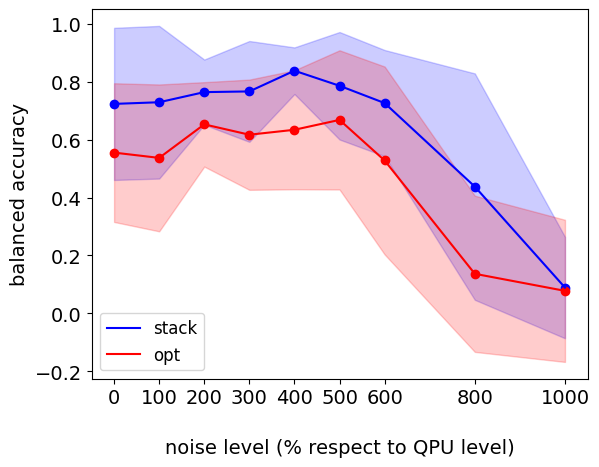}  
  \caption{Balanced accuracy}
  \label{fig:sub-second}
\end{subfigure}
\caption{Scaling of performances with 5 training samples, 10-atom simulations.}
\label{fig:scaling N=5}
\end{figure}

\begin{figure}[ht]
\begin{subfigure}{.5\textwidth}
  \centering
  \includegraphics[width=1.0\linewidth]{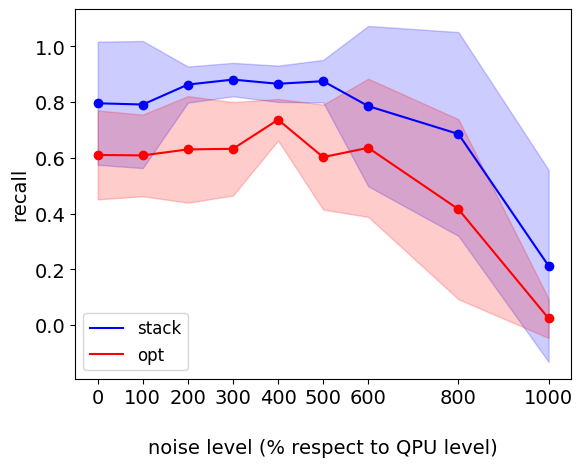}  
  \caption{Recall}
  \label{fig:sub-first}
\end{subfigure}
\begin{subfigure}{.5\textwidth}
  \centering
  \includegraphics[width=1.0\linewidth]{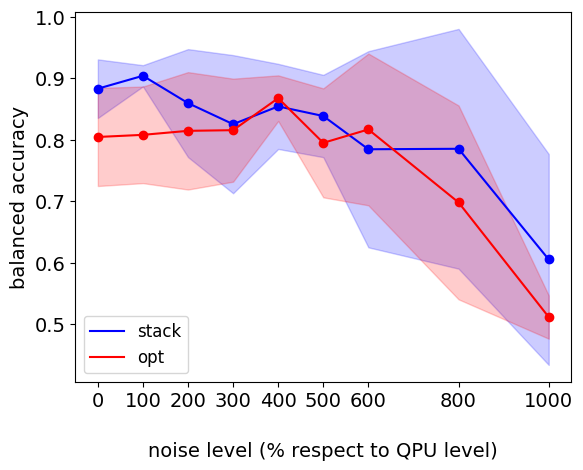}  
  \caption{Balanced accuracy}
  \label{fig:sub-second}
\end{subfigure}
\caption{Scaling of performances with 7 training samples, 14-atom simulations.}
\label{fig:scaling N=7}
\end{figure}

The scaling analysis is done by fixing the number of atoms and varying the noise level. The obtained results are shown in Fig.\ref{fig:scaling N=5} Fig \ref{fig:scaling N=7} for 5 and 7 training samples (10 and 14 atoms) respectively.
The noise levels shown in the x-axis are reported as a percentage of those used with respect to Fig. \ref{fig:trendline_recall_new} and \ref{fig:trendline_bal_acc_new} (and present on the actual \ac{qpu}). This allows us to see how performance varies from ideal simulations to simulations with $1000\%$ of the noise levels compared to what is claimed on the \ac{qpu}. What we can see in general is a bell-shaped trend, i.e. higher performance and lower errors are obtained when some noise is present compared to its total absence. Therefore, there is evidence of performance enhancement due to noise.
This is a new phenomenon in the context of quantum machine learning, but known for some time in other contexts.
For example, there are cases in which the presence of noise has a beneficial effect, such as quantum transport over complex networks and quantum maze escapes \cite{Caruso_Datta_09,Caruso_Crespi_quantum15, Caruso2016_qc}.

Summarizing, an analysis of the results in Figures \ref{fig:recall scaling} and \ref{fig:balacc scaling} shows that the best models in terms of maximum performance and minimum error are the QUBO SVM models in a stacked configuration. In particular, the version trained with noisy simulations seems to offer better results and smaller error bars. This is supported by what is shown in Fig. \ref{fig:scaling N=5} and \ref{fig:scaling N=7}.
In addition, the models trained with \ac{qpu} show no particular drop in performance, although the real noise is more complex than the noise models implemented on quantum libraries. This allows us to test the performance of the model successfully by exploring regimes that we are unlikely to access through simulations, noting how the standard deviations remain small as the number of atoms increases.

\subsection{Computational complexity}
In this section we aim to analyze time-qubit complexity of the introduced \ac{qubo} \ac{svm} model, and compare it with the commonly used classical \ac{svm} models.
We have to distinguish three different phases in which we will derive the complexity:
\begin{enumerate}
    \item  time required to convert the algorithm into a \ac{qubo} problem
    \item  time required to embed the \ac{qubo} problem on the quantum register
    \item theoretical time complexity of QAA. 
\end{enumerate}
The former can be derived from Eq. \eqref{qubo complexity} and it is estimated by $\mathcal{O}(N^2 \times K^2 \times N_{features})$ operations if a linear kernel is exploited.
Regarding the number of operations required by the embedding, it is constant, since we specify a maximum number of iterations in the COBYLA optimizer. Therefore, there is no real scaling. Morover, the number of features $N_{features}$ of the dataset is fixed and can be reduced via dimensionality reduction techniques.

Finally, although the theoretical time complexity of \ac{qaa} is $\mathcal{O}(1/\Delta^2)$, where $\Delta$ is the minimum eigenvalue gap
between the ground state and the first excited state of
the Hamiltonian $H$. In any case, since in the current neutral atom \ac{qpus} the maximum sequence duration is upper bounded, we consider it to be constant and equal to this limit. Therefore, even in this case, there is no real scaling to take into account.
Therefore, the time complexity of the proposed quantum algorithm is $\mathcal{O}(N^2 \times K^2)$.
In contrast, the time complexity of \ac{svm} is $\mathcal{O}(N^3)$, although recently algorithms have been proposed to lower the complexity to $\mathcal{O}(N^\gamma)$, with $2<\gamma<3$.
Finally, regarding the qubit footprint, our algorithm shows a $\mathcal{O}(N \times K)$ qubit scaling. 
However, since $K$ is a fixed hyperparameter, the complexity reduces to $\mathcal{O}(N^2)$ and $\mathcal{O}(N)$ for time complexity and qubit footprint respectively.

Therefore, the algorithms proposed in this manuscript have: better scaling than their classical counterparts ($N_{QUBO}^{2}$ vs $N_{SVM}^{2<\gamma<3}$); better (or at best) performance than classical models and lower error bars; the possibility of being implementable on atom-neutral \ac{nisq} devices; little use of the \ac{qpu}; and the privacy of the data is preserved because only atomic coordinates and laser parameters reach the \ac{qpu}, the model test is done locally because only training requires a \ac{qpu}.

\section{Conclusion}\label{sec:conclusion}
We have shown a model of \ac{svm} whose training is reformulated through in a QUBO problem. These kinds of problems are naturally implementable on neutral atom devices using analog processing, that is, performing only global laser operations on the atoms. The linear classifier model can be easily used in the presence of nonlinear data in a variety of ways: using a nonlinear kernel, via a kernel PCA, or via feature discretization.
The model was trained and tested on a very unbalanced \ac{ccf} detection dataset. This required first the application of all the techniques used in the case of highly unbalanced datasets: namely, the use of resampling techniques and the choice of appropriate metrics (other than accuracy, which is definitely obsolete in these cases). The model we considered was tested in several variants considering also ensembling strategies. For each strategy, the trained model was tested by both ideal and noisy simulations. To provide robust estimates with associated measures of error, each configuration was trained and tested on 10 different splits of the dataset finally calculating mean and standard deviation of the chosen metrics. 
In addition, in order to numerically study the scalability of performance, these analyses were performed by running ideal and non-ideal simulations with training sets of 4, 5, 6, 7 and 8 samples, using quantum registers of 8, 10, 12, 14 and 16 atoms, respectively. Finally, the same analysis was then carried out using a real \ac{qpu} with neutral atoms. In addition to the configurations already used and listed above, quantum registers of 18, 20, 22 and 24 atoms were also used; corresponding to training sets of 9, 10, 11 and 12 samples. All QUBO SVM models were compared with the main binary classifier (classical) models in the literature, resulting in performance comparisons.  Finally, the same analysis was then carried out using a real \ac{qpu} with neutral atoms. In addition to the configurations already used and listed above, quantum registers of 18, 20, 22 and 24 atoms were also used; corresponding to training sets of 9, 10, 11 and 12 samples. We have faced that our proposed models have performances that do not deviate much from the classical models and are compatible with them within the error bars. However, a stacked ensemble variant we proposed seems to suggest superior performance and smaller error bars, both in terms of recall and balanced accuracy. In particular, focusing only on this type of model and analyzing its scaling of both the ideal, noisy and QPU versions, some interesting conclusions can be drawn. The first is that the performance seems to increase as the number of atoms (and thus training samples) increases Based on this, we believe that by using a few hundred training samples, it may be possible to achieve superior performance to classical models. This means that the model could lead to higher metric values and lower associated errors. This is important because it motivates interest in investigating its performance by increasing the number of atoms even more. For this reason, it would be interesting to test it on a neutral-atom \ac{qpu} with 100-200 qubits, in order to verify the actual robustness under real conditions. Another interesting consideration concerns the fact that both the noisy and \ac{qpu} model, in addition to having almost identical performances as the ideal one, shows slightly lower error. To confirm this observation, we performed and compared simulations at different noise levels, showing improvements in the presence of noise compared to the ideal case.
This could be related to the fact that quantum noise in \ac{qaa} can help to find better solutions, i.e. global minima of the cost functions. Finally, here there are some remarks on the QUBO SVM model.
It is trained adiabatically via analog quantum computing and the data is encoded via the $Q$ matrix. This avoids using the various types of encoding used in digital computation that often result in quantum circuits that are too long to execute, hence not feasible on \ac{nisq} machines. Therefore, the increase in the number of qubits is not a problem in this respect. In fact, the part of the circuit responsible for encoding, which is absent here, grows in depth with the number of qubits. Moreover, the fact that the model is trained only on the \ac{qpu} is advantageous because it reduces the use of the \ac{qpu} and the associated costs. Since only the training is quantum, the model can be saved and used later or multiple times at the same cost. Such a model is suitable also for contexts where data privacy is a crucial factor: in fact, the training data are never disclosed, but are used locally to compute the $Q$ matrix. In turn, the $Q$ matrix is reproduced as closely as possible from the Hamiltonian of the system through device optimization. Therefore, even in case of data leakage during the connection with the \ac{qpu}, only the coordinates of the atoms and the intensity and detuning values would be disclosed, or at the limit, the results of the measurement without any informations on the data used for the training. The analysis of the test data is again performed locally. This study motivates the development and use of model training protocols based on analog processing in real-world scenarios, given the low requirements compared to gate-based versions, in anticipation of fault-tolerant quantum computers. Others possible developements could include the unsupervised version of \ac{svm} known as one-class \ac{svm}, a version of \ac{svm} with a modified loss so as to assign different weights to misclassifications in cases where the datasets are not balanced and the use of counterdiabatic drivings, which are useful to prevent excitation while preventing the adiabatic protocol from becoming too long.
Although noise does not seem to be a problem in the explored regimes, we will study what benefits we can get from logical qubits, and how they can be moved to change the topology of the quantum register, further improving our results.

\section*{Acknowledgements}
We acknowledge the CINECA award under the ISCRA initiative for the availability of high performance computing resources and for their support. Finally, we thank Pasqal for their constant support and for running the simulations necessary to test our protocol on their QPU.


%
\printbibliography

\section{Methods}

\subsection{SVM}\label{appendix:SVM}
The classical \ac{svm} algorithm involves the resolution of the following quadratic programming problem:
\begin{equation}\label{svm training}
    \text{minimize }  \left[ \frac{1}{2} \sum_{n,m} \alpha_n \alpha_m y_n y_m k(x_n, x_m) - \sum_{n} \alpha_n \right]
\end{equation}
\begin{equation}\label{first constraint}
    \text{subject to   } 0 \leq \alpha_n \leq C,
\end{equation}
\begin{equation}\label{second constraint}
    \text{and   } \sum_n \alpha_n y_n = 0,
\end{equation}
with $C$ regularization parameter, $k(\cdot)$ kernel function and $\alpha$ coefficients.

Each $\alpha_n$ coefficient defines a decision boundary of dimension $d-1$ separating $\mathbb{R}^{d}$ in two regions corresponding to the two classes.
Given a decision boundary defined by $\alpha_n$, a prediction for a generic sample $x_j$ can be made evaluating the decision function:
\begin{equation}\label{decision function}
    f(x_j) = \sum_n \alpha_n y_n k(x_n, x_j) + b, 
\end{equation}
where the bias term $b$ is given by
\begin{equation}\label{bias}
   b = \frac{ \sum_n \alpha_n (C -  \alpha_n)[y_n - \sum_m \alpha_m y_m k(x_n, x_m)]}
   { \sum_n \alpha_n (C -  \alpha_n)}.
\end{equation}
Since the decision function $f$ represents the signed distance of a sample from the decision boundary, the predicted class label is obtained with 
\begin{equation}
    \hat{y_j} = sgn(f(x_j)).
\end{equation}
It can be seen that the complexity of the problem does not increase with the number of features ($d$), since the data appear only as values of the kernel function $k(\cdot)$.
This fact is known as the "kernel trick" and can be exploited to map the data into higher dimensional spaces where, hopefully, the data separates more easily.

\subsection{QUBO Support Vector Classifier}\label{appendix:QUBO SVM}
Once the \ac{svm} model and the \ac{qubo} problem have been introduced, it can be seen that it is possible to reformulate the training process of a \ac{svm} model
by rewriting it as a \ac{qubo} problem that will be solved by \ac{qaa}. Such a model will be called \ac{qubo} \ac{svm}.

However, before rewriting the cost function, we need to encode the data since the quantum computer used in analog way to perform \ac{qaa} can produce only binary solutions.
We encode the data defining
\begin{equation}\label{encoding}
    \alpha_n = \sum_{k=0}^{K-1} B^k a_{Kn+k},
\end{equation}
where $ a_{Kn+k} \in \{ 0,1\}$, $K$ is the number of binary variables used to encode $\alpha_n$ and $B$ is the base used for the encoding. Introducing a multiplier $\xi$ we can modify the \eqref{svm training} in the following cost function:
\begin{equation}\label{qubo energy}
\begin{split}
        E = \frac{1}{2} \sum_{n,m,k,j} a_{Kn+k} a_{Km+j} B^{k+j}y_n y_m k(x_n, x_m) - \\
    \sum_{n,k} B^k a_{Kn+k} + \frac{\xi}{2}( \sum_{n.k}B^k a_{Kn+k y_n})^2 
\end{split}
\end{equation}
\begin{equation}\label{qubo complexity}
     = \sum_{n,m=0}^{N-1} \sum_{k,j=0}^{K-1} 
     a_{Kn+k} Q_{Kn+k, Km+j} a_{Km+j}, 
\end{equation}
with 
\begin{equation}\label{qubo matrix}
    Q_{Kn+k, Km+j} = \frac{1}{2}B^{k+j} y_n y_m (k(x_n, x_m) + \xi) - \delta_{nm}\delta_{kj}B^k
\end{equation}
known as \ac{qubo} matrix of size $KN \times KN$ and $N$ size of the training set.
We can notice that the constraint \eqref{first constraint} is included in the encoding \eqref{encoding} since for the maximum for $\alpha_n$ is given by
\begin{equation}
    C = \sum_{k=1}^{K} B^k
\end{equation}
and $\alpha_n \geq 0$ by construction.
Also the constraint \eqref{second constraint} is included in \eqref{qubo energy} through the multiplier $\xi$. In out manuscript this we choose a value near to one, namely $\xi=0.5$.

\subsection{Ensemble Methods}\label{appendix:ensemble}
\begin{figure}[ht]
\begin{subfigure}{.85\textwidth}
  \centering
  \includegraphics[width=1.\linewidth]{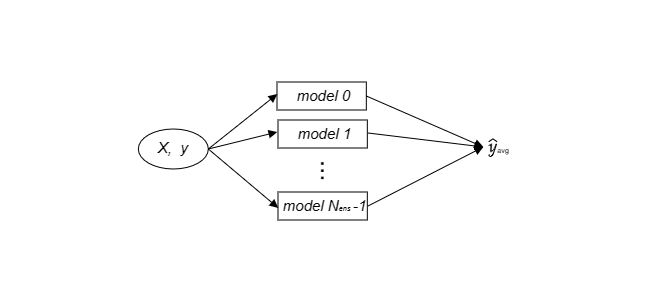}  
  \caption{}
  \label{fig:averaging}
\end{subfigure}
\begin{subfigure}{.85\textwidth}
  \centering
  \includegraphics[width=1.\linewidth]{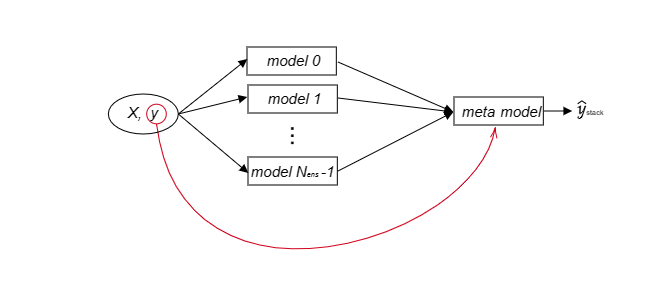}  
  \caption{}
  \label{fig:stacking}
\end{subfigure}
\caption{(a) Outline of the ensembling approach of average voting type, the final predictions are the average of the predictions of the ensemble of classifiers. (b) Outline of the stacking type ensembling approach, the final predictions are the predictions of the meta model.}
\label{fig:ensemble}
\end{figure}

Ensembling is a powerful and useful method in machine learning. It is based on the intuition that combining the predictions of many weak learners may lead to a more powerful learner. In fact, weak learners are by definition easy to train-models that can performs slightly better than random guessing. Conversely, strong learners are powerful models that require lot of training and they are usually hard to obtain. Ensemble learning is a group of techniques whose goal is to obtain strong learners by combining together many weak learners. Among the advantages there are: increased performances, robustness and reliability.
Two of the most famous ensemble techniques are average voting and stacking.
In the first, as shown in figure \ref{fig:averaging}, all the predictions of the $N_{ens}$ weak learners, $\hat{y}^{weak}$, are averaged:
\begin{equation}
    \hat{y}_{avg} = \frac{1}{N_{ens}} \sum_{i=0}^{N_{ens}-1} 
    (\hat{y}^{weak})_i.
\end{equation}
Instead, in stacking, two different layers are defined: there is a first layer made by different learners that are trained in the usual way. Then, as shown in figure \ref{fig:stacking}, the second layer consists of a meta model trained using as input the joined predictions of the first layer and as labels the original ground truths:
\begin{equation}
    \hat{y}_{stack} =  f_d^{meta}([(\hat{y}^{weak})_0, (\hat{y}^{weak})_1, \dots, (\hat{y}^{weak})_{N_{ens}-1}]),
\end{equation}
where $f_d^{meta}$ is the decision function of the meta model.
In stacking the base learners in the first layer are used as a feature extractor.

\subsection{Models}\label{appendix:models}
Table \ref{tab:models} shows all the models used in this manuscript. The first column shows the type of model, the second gives a brief description of the model to make it easier to distinguish the various versions. Finally, the last column contains the acronym that refers to the model and will be placed in the graphs.
\begin{table}
    \centering
    \begin{tabular}{|p{0.15\textwidth} |p{0.65\textwidth}| p{0.2\textwidth}|}\hline
         Model& Description &Name in the graph\\ \hline 
         KNN& Number of neighbors equal to 3 &KNN\\ \hline 
         Random Forest&  &Random Forest\\ \hline 
 Decision Tree& &Decision Tree\\ \hline 
 Naive Bayes& &Naive Bayes\\ \hline 
 Logistic Regression& &Logistic Regression\\ \hline 
 SVM&Linear Kernel &SVM Lin\\ \hline 
 SVM& RBF Kernel&SVM RBF\\ \hline 
 QUBO SVM& Trained with ideal simulation&QUBO SVM i\\ \hline 
 QUBO SVM& Trained with ideal simulation, the number of models to be aggregated is optimized with respect to recall&QUBO SVM i opt\\ \hline 
 QUBO SVM& Trained with noise simulation,  final state sampled with 1000 shots &QUBO SVM N \\ \hline 
 QUBO SVM& Trained with noise simulation, the number of models to be aggregated is optimized with respect to recall, final state sampled with 1000 shots &QUBO SVM N opt \\ \hline 
 QUBO SVM& Trained with noise simulation,  final state sampled with 500 shots &QUBO SVM 500\\ \hline 
 QUBO SVM&Trained with noise simulation, the number of models to be aggregated is optimized with respect to recall, final state sampled with 500 shots  &QUBO SVM N opt 500\\ \hline 
 QUBO SVM& Trained with noise simulation,  final state sampled with 100 shots&QUBO SVM N 100\\ \hline
 QUBO SVM&Trained with noise simulation, the number of models to be aggregated is optimized with respect to recall, final state sampled with 100 shots  &QUBO SVM N opt 100\\ \hline 
 QUBO SVM& Trained with a \ac{qpu} simulation,  final state sampled with less than 100 shots & QUBO SVM opt QPU\\ \hline
  Stacked ensemble&QUBO SVM trained via ideal simulations, used as meta model in a stacked configuration where the models in the first layer are: Naive Bayes, Random Forest, Logistic
Regression and KNN  &QUBO SVM i Stack\\ \hline 
Stacked ensemble&QUBO SVM trained via noise simulations, used as meta model in a stacked configuration where the models in the first layer are: Naive Bayes, Random Forest, Logistic
Regression and KNN   &QUBO SVM N Stack\\ \hline
Stacked ensemble& QUBO SVM trained with a simulation run on the \ac{qpu}, used as meta model in a stacked configuration where the models in the first layer are: Naive Bayes, Random Forest, Logistic
Regression and KNN & QUBO SVM Stack QPU\\ \hline
    \end{tabular}
    \caption{Model used for the comparative analysis. In the second column there are the detailed descriptions of all the models. In the third one there are the names that identifies the models in the graphs.}
    \label{tab:models}
\end{table}

\section{Results}\label{appendix:results}

\end{document}